\documentclass[sigconf,authorversion]{acmart}

\AtBeginDocument{%
  }

\copyrightyear{2023}
\acmYear{2023}
\setcopyright{acmlicensed}\acmConference[ARES 2023]{The 18th International Conference on Availability, Reliability and Security}{August 29-September 1, 2023}{Benevento, Italy}
\acmBooktitle{The 18th International Conference on Availability, Reliability and Security (ARES 2023), August 29-September 1, 2023, Benevento, Italy}
\acmPrice{15.00}
\acmDOI{10.1145/3600160.3605006}
\acmISBN{979-8-4007-0772-8/23/08}

\begin{document}

\title{A Study of Different Awareness Campaigns in a Company}

\author{Laura Gamisch}
\email{l.gamisch@campus.lmu.de}
\affiliation{%
  \institution{Ludwig-Maximilians-Universit\"at M\"unchen}
  \city{Munich}
  \country{Germany}
}

\author{Daniela P\"ohn}
\orcid{0000-0002-6373-3637}
\email{daniela.poehn@unibw.de}
\affiliation{%
  \institution{Universit\"at der Bundeswehr M\"unchen, RI CODE}
  \city{Neubiberg}
  \country{Germany}
}
\renewcommand{\shortauthors}{Gamisch and P\"ohn}

\begin{abstract}
Phishing is a major cyber threat to organizations that can cause financial and reputational damage, threatening their existence. The technical measures against phishing should be complemented by awareness training for employees. However, there is little validation of awareness measures. Consequently, organizations have an additional burden when integrating awareness training, as there is no consensus on which method brings the best success. This paper examines how awareness concepts can be successfully implemented and validated. For this purpose, various factors, such as requirements and possible combinations of methods, are taken into account in our case study at a small- and medium-sized enterprise (SME). To measure success, phishing exercises are conducted. The study suggests that pleasant campaigns result in better performance in the simulated phishing exercise. In addition, significant improvements and differences in the target groups could be observed. The implementation of awareness training with integrated key performance indicators can be used as a basis for other organizations.
\end{abstract}



\maketitle

\section{Introduction}
\label{sec:introduction}

Since the first ARPANET network mail in 1971, a series of Request for Comments (RFCs) have standardized the current form of the email. As the number of email users continues to grow each year, email has evolved into a core communication method and is now an integral part of daily Internet life. The number of global email users was 4 billion in 2020 and is estimated to grow to 4.6 billion in 2025~\cite{statista}. In 2020, approximately 306 billion emails were sent and received every day, but not all emails are well intended. Malicious emails can compromise digital accounts, devices, and, consequently, organizations. According to Verizon's Data Breach Report 2022~\cite{verizon}, 82\% of breaches involve the human element, with phishing being the second-most common entry point into an organization. The subsequent incidents have various impacts ranging from financial to reputational damage~\cite{10.1007/978-981-16-4884-7_3}, as these attacks are often one step towards the goal within the cyber kill chain~\cite{8551383,9265953,9789803}. For example, MITRE Adversarial Tactics, Techniques, and Common Knowledge (ATT\&CK) assigns spear-phishing the IDs T1566.001-003 (with attachment, via a link, or via a service) within the phase initial access~\cite{mitreid}.

While technical countermeasures help to prevent and detect several attacks, some malicious emails might not be noticed by them and, consequently, be delivered to employees. Hence, it is then up to the employees to act in a desirable way, such as by deleting those emails or notifying the internal support team. This is only possible when they know about the social engineering attack of phishing, which targets the human element~\cite{10051531}, and the consequences of successful attacks. In order to help them identify phishing emails and react accordingly, awareness training can be provided. Although the literature gives a broad overview of training campaigns (see, for example, \cite{10.1145/3299815.3314437}), there is no consensus on which type of awareness training is best for all or groups of employees based on their characteristics~\cite{9753912}.

In order to close the gap, we design and implement three different awareness training methods in an SME and verify the results based on simulated phishing exercises. Additional surveys provide insights into the requirements for awareness training. Consequently, the contribution focuses on 1) KPIs for phishing training, 2) requirements and effectiveness of awareness training campaigns, and 3) feedback and attendance of the employees.

The remainder of the paper is as follows: First, we outline our methodology in Section~\ref{sec:methodology}. In Section~\ref{sec:prestudy}, we describe the study design and the results of the pre-study. Following in Section~\ref{sec:concept}, we describe the concept of the phishing study, consisting of simulated phishing exercises, awareness campaigns, and the evaluation of its effectiveness. The results are shown in Section~\ref{sec:study}, followed by a discussion. Section~\ref{sec:relatedwork} contrasts our study with related work. Finally, we conclude the paper and state future work.

\section{Methodology}
\label{sec:methodology}

To compare the different types of training campaigns, the process includes two surveys and two phishing exercises.

\begin{description}
\item[Process:] All stages were carried out within six months.
\begin{enumerate}
\item Survey 1: Pre-study to compare different awareness campaigns and gain requirements from a wider range of participants.
\item Simulated phishing exercise 1: Simulated phishing exercise based on a real phishing email in the SME.
\item Awareness training: Three different awareness training sessions, i.\,e. circular (group 1), quiz (group 2), and online seminar (group 3). In addition, every participant received a detailed handout. 
\item Simulated phishing exercise 2: Simulated phishing exercise based on another real phishing email.
\item Survey 2: Post-study to assess the different awareness training campaigns in the SME.
\end{enumerate}
\item[Recruitment:] We had two different participant groups (A: two universities, B: SME). Groups 1-3, which are mentioned in stage 3, are only part of group B. Neither compensations nor prizes were given. 
\begin{itemize}
\item Survey 1 (stage 1): 58 students and employees from two universities and the SME, recruited by mailing lists and other means of communication over the course of three weeks (group A and B).
\item  Phishing Study and Survey 2 (stages 2-5): SME with around 100 employees in the field of IT service sector (only group B).
\end{itemize}
\item[Survey design:] The survey was multifold: 
\begin{enumerate}
\item pseudonymization, 
\item technical knowledge, 
\item detection of phishing, 
\item awareness training campaigns (knowledge, rating, and requirements), 
\item demographic information.
\end{enumerate}
\item[Ethics and data protection:] The study was in compliance with the ethical board of the university and the SME, and obeyed the required data protection measures according to the General Data Protection Regulation (GDPR). In addition, a contract with the SME was signed.
\item[Limitations:] The number of participants in general and from the SME in the first survey is limited. Only three employees participated in both surveys. Hence, no direct comparison was made. In addition, the results only stand for the SME and might differ from organization to organization.
\end{description}

\section{Pre-Study}
\label{sec:prestudy}

The goal of the pre-study was to analyze which awareness training campaigns against phishing are known and which requirements the participants have for those training methods. In addition, it was asked how enjoyable and effective these training campaigns are perceived. We first outline the study design (see Section~\ref{sec:design}), before the results are discussed in Section~\ref{sec:results}.

\subsection{Study Design}
\label{sec:design}

The study design follows the summary described in Section~\ref{sec:methodology}. Parts 2-4 are explained in more detail. To estimate the technical knowledge of the participants (part 2), they had to answer questions related to the scale of their technical-savvy, their recognition of phishing emails, and their general knowledge about phishing. Following this, the participants were asked to name the characteristics of phishing emails (part 3). Next, the participants could state up to seven awareness training campaigns (part 4). Each training was then graded according to its effectiveness and joyfulness. Additionally, combinations and requirements for awareness training were requested.

The main focus of the survey was to acquire the knowledge of the participants instead of letting them choose from pre-defined answers. Hence, text boxes were provided. Although the answers were not influenced, this made the evaluation more difficult. Twelve of the participants were able to name seven characteristics of phishing emails, which was also the pre-defined maximum of possible answers. More fields might have led to further stated characteristics.

\subsection{Study Results}
\label{sec:results}

58 persons participated in the survey. The participants tended to be technical-savvy and comparably young (18-29 years: 27, 30-49 years: 17, 50-69 years: 9). More females (34) took part than males (18). The proportion of students (22) to employees (29) was almost even.

The stated characteristics of phishing emails include the sender address (47), followed by the language (35), links (34), specific content (25), urgency (20), request for data (19), salutation (28), and layout (17). Additionally, attachment (8), subject (6), spam filter (4), technical details such as error messages (3), receiving address (2), mass email (1), and gut feeling (1) were named. On average, the participants knew four characteristics. Around half of the participants were able to mention more than four characteristics. The self-evaluation of the technical knowledge (part 2) correlates with the stated characteristics.

The participants identified 74 awareness campaigns. Eleven participants named general sensitization without going into detail and several participants did not know about awareness at all. The awareness campaigns were then evaluated by the participants and grouped into categories by the authors, as shown in Table~\ref{tab:pre}.

\begin{table}[!htpb]
	\centering
	\caption{Stated awareness trainings with their evaluation}\label{tab:pre}
	\begin{tabular}{lccc}
		\toprule
		\textbf{Awareness training} & \textbf{\#} & \textbf{Effective} & \textbf{Joy} \\ \midrule
		Training/webinar/seminar & 27 & 56\% & 48\% \\ 
		Phishing exercise & 15 & 93\% & 40\%\\ 
		Frequent information & 8 & 50\% & 75\% \\ 
		Newsletter/circular & 8 & 38\% & 38\% \\ 
		Automatic aids & 5 & 40\% & 80\% \\ 
		Activate participants & 4 & 75\% & 50\%\\ 
		Handout & 3 & 100\% & 100\% \\ 
		Posters & 2 & 50\% & 50\% \\ 
		Others & 2 & 100\% & 0\% \\ 
		\bottomrule
	\end{tabular}
\end{table}

\section{Concept of the Phishing Study}
\label{sec:concept}

This section describes the concept of the main study, consisting of the simulated phishing exercises (see Section~\ref{sec:concept-training}) and the three awareness campaigns (see Section~\ref{sec:concept-awareness}), and the survey to evaluate the effectiveness (see Section~\ref{sec:evaluation-effect}).

\subsection{Simulated Phishing Exercises}
\label{sec:concept-training}

The simulated phishing exercise should provide KPIs of phishing awareness and a comparison between the different awareness campaigns, as explained in Section~\ref{sec:concept-awareness}. Hence, the three groups are divided evenly. The phishing exercise is performed twice: once before and once after the awareness campaigns. Considered KPIs are click rate (``Clicked Link''), error rate (``Submitted Data''), reporting rate, and resilience factor. The resilience factor represents the ratio between the error rate and reporting rate and is specified as follows (1). If the reporting rate is smaller than the error rate, then (2) applies.

\begin{align}
\text{resilience rate} &= \frac{\text{reporting rate}}{\text{error rate}}\\
\text{resilience rate} &=  - \frac{\text{error rate}}{\text{reporting rate}}
\end{align}

Different phishing tools exist, for example, Evilginx2, Modlishka, GoPhish, King-Phisher, Social Engineering Toolkit, and Microsoft Simulation Training. Based on the features and last date of update, the tool GoPhish~\cite{gophish,10.1007/978-3-030-65610-2_9} is chosen. It is set up on a hardened and securely configured virtual Linux machine with an external Internet protocol (IP) address. For both simulated phishing exercises, domains spelled similarly to the original domain are utilized. Secure sockets layer (SSL) certificates are issued to these domains. In addition, the sender policy framework (SPF) entry is registered for both domain name systems (DNS). For each domain, a sending profile for the underlying email service is added. The phishing emails used for the exercise are taken from public repositories and adapted for the SME (regarding language, salutation, uniform resource locator (URL) obfuscation, request to click on the link, etc.). Each link to the phishing website contains a tracking parameter to control the KPIs. Alternatively, received phishing emails could be traced as KPIs using the mentioned parameter in the form of an empty image appended to the email. The emails are sent after the lunch break in a drop-wise manner to reduce the impact of chatter and the likelihood of account closure. The emails are tested before the lunch break with selected people involved.

\begin{figure}[!htbp]
	\centering
	\includegraphics[width=0.5\textwidth]{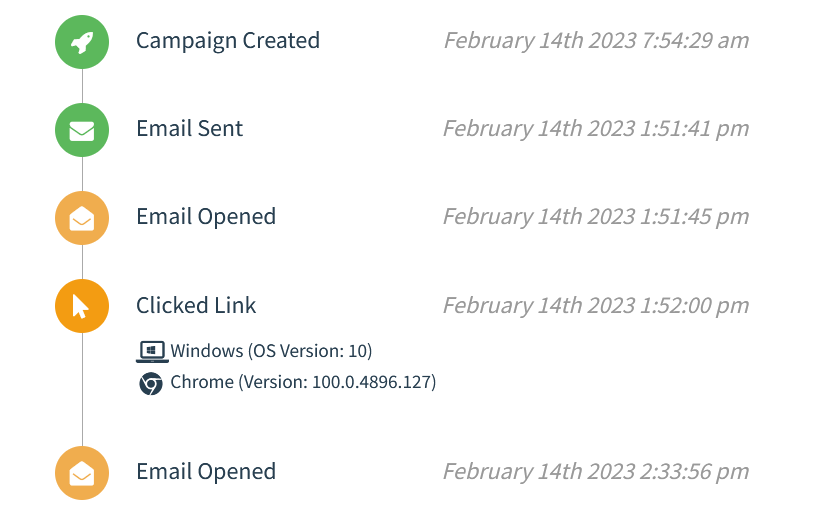}
	\caption{``E-Mail Opened'' after ``Clicked Link'' event caused by spam filter}
	\Description[``E-Mail Opened'' after ``Clicked Link'' event caused by spam filter]{The event status``E-Mail Opened'' after the status ``Clicked Link'' is caused by the applied spam filter} 
	\label{fig:gophish_events} 
\end{figure}

Publicly accessible login pages of the SME could be used for phishing attacks and are consequently copied for the exercises. The URL also contains a unique parameter. Based on this parameter, the KPIs are determined as multiple actions from one user are only counted once. No input data is stored and the employees are forwarded to the real websites. The templates of GoPhish are adapted to respect privacy and data protection. When evaluating the KPIs regarding ``Clicked Link'' and ``E-Mail Opened'', the result may be distorted. We assume that this could be the case due to the spam detection mechanisms of the email provider. If the user has opened the email but not clicked on the URL, the result would be positive for clicking if the spam filter has preprocessed the given URL (see Figure~\ref{fig:gophish_events}). The measured value can be reviewed and adjusted manually.

\subsection{Awareness Campaigns}
\label{sec:concept-awareness}

In order to compare the different awareness training campaigns, the employees are randomly divided into the following three groups. In addition, a handout about phishing is integrated into the information security management system (ISMS).
\begin{itemize}
\item Group 1: Circular email and reference to the handout.
\item Group 2: Email, quiz and confirmation of participation.
\item Group 3: Two emails, online training and confirmation of participation.
\end{itemize}

The participants in group 1 receive a circular email with the most important information about phishing, such as the five main characteristics of phishing emails, and a reference to the handout. It cannot be reproduced if the participants read the handout. This group is used as a baseline, as such an email was already sent in June 2022.

Based on the pre-study, theoretical and practical parts should be combined for awareness training to convey the theoretical information and activate the employees. As an online quiz and a training session include active parts, both forms are chosen. The participants of group 2 receive a similar email to the first group, but have to solve an online quiz with the classification into normal or phishing emails of eight exemplary emails and a final question in another tool to confirm their participation.

The participants in group 3 receive a calendar invitation to an online training session via email. Brief information about the training is provided in an additional email. The text of the email is shorter, as most information is conveyed in the training. The training consists of (a) an introduction including the amount of damage and importance of data; (b) retrospection of the first phishing simulation as motivation; (c) technical measures and the importance of awareness; (d) characteristics of phishing emails with a focus on the five most important characteristics; (e) desired reactions to phishing emails; and (f) a summary of the content. During the presentation, the participants are asked interactive questions (between (a) and (b) or (e) and (f) respectively). The participants have to confirm their participation in another tool by answering one question. As the online training is recorded, the participants have the option to see it asynchronously.

\subsection{Evaluation of the Effectiveness}
\label{sec:evaluation-effect}

To evaluate the different awareness training campaigns, not only the KPIs, described in Section~\ref{sec:concept-training}, are determined, but also how effective and joyful the participants rate the individual methods. In order to receive this input, another survey is created. The survey has a similar structure as the one of the pre-study to compare the results. In contrast, no text boxes but scales from 0 to 100 are applied for rating the effectiveness and joyfulness of training campaigns and which methods the participants prefer. Moreover, further questions related to future topics, the training in which particpants took part, the information in the ISMS handbook, and the observation of other training campaigns are asked.

As the participants of the post-study also participated in the awareness campaigns, their answers might be biased toward their participation. Further demographic factors and technical-savvy were not considered and should be part of another study.

\section{Results of the Phishing Study}
\label{sec:study}

The phishing study was conducted between December 2022 and March 2023, with preparations beforehand. In this section, we describe the results of the first simulated phishing exercise (see Section~\ref{sec:first}), the three awareness campaigns (see Section~\ref{sec:campaigns}), the second simulated phishing exercise (see Section~\ref{sec:second}), the comparison of both exercises (see Section~\ref{sec:comparison}), and the post-study (see Section~\ref{sec:post}).

\subsection{First Phishing Exercise}
\label{sec:first}

After the emails were sent to the employees, nine persons asked the internal support team whether the email was an actual phishing email. These persons were informed about the simulated phishing exercise with the request not to share this information with colleagues. As five persons belong to group 2, it might be that this group has more knowledge about phishing. One employee noticed the phishing attempt due to the URL parameters, which are not clearly visible in the email.

Overall, 26\% of the participants clicked the link, 17\% entered their data, and at least 6\% loaded the tracking picture. No significant variance between the groups could be observed, except that more participants in group 2 clicked the link. Although 10 males and 10 females clicked the link, more female participants submitted their data (eight females and five males). As a result, the reporting rate is 11\%, the error rate is 17\%, and the resilience factor is -1.5. Such rates were expected, as no previous phishing awareness training besides one circular (six months ago) was conducted.

\subsection{Awareness Campaigns}
\label{sec:campaigns}

Before the three awareness campaigns, the result of the first simulated phishing exercise was presented to the employees in their monthly meetings. The following queries of employees related to the exact email, if one entered their data, and how to react to such an email. In addition, more phishing emails were reported to the internal support team. Afterward, the three awareness campaigns were conducted. Groups 2 and 3 had to answer one question in an online quiz to verify their participation. Both groups consisted of 28 participants. 38 participants (15 of group 2  and 23 of group 3) answered the question correctly, 6 (5/1) incorrectly, and 12 (8/4) did not answer it at all.

\subsection{Second Phishing Exercise}
\label{sec:second}

The selected phishing email for the second exercise, which employs the factor of fear of consequences, was sent to 83 employees after lunchtime. The procedure thereby applies the concept of the first phishing exercise. 19 participants informed the internal support team and were enlightened about the exercise. The number of notices varies between the groups (group 1: 3; group 2: 6; and group 3: 10). As the information about the phishing exercises and their evaluation was transparent, the concerns stated by Volkamer et al.~\cite{volkamer} were falsified.

Except for two employees, no one clicked the link without also entering their data. The error rate is 5\% on average and, thereby, improved in contrast to the first exercise. Reporting rate, error rate, and resilience factor are evaluated per group as displayed in Table~\ref{tab:comparison-group}.

\begin{table}[!htpb]
	\centering
	\caption{KPIs of the survey groups for the second exercise}\label{tab:comparison-group}
	\begin{tabular}{lccc}
		\toprule
		\textbf{Group} & \textbf{Reporting} & \textbf{Error} & \textbf{Resilience} \\ \midrule
		Group 1 (circular) & 11\% & 7\% & 1.6  \\ 
		Group 2 (quiz) & 21\% & 4\% & 5.3  \\ 
		Group 3 (training) & 36\% & 4\% & 9.0  \\ 
		\midrule
		Over all groups & 23\% & 5\% & 4.6  \\ 
		\bottomrule
	\end{tabular}
\end{table}

\subsection{Comparison of the Phishing Exercises}
\label{sec:comparison}

Both simulated phishing exercises show similarities, but also differences. The average reporting rate of the first phishing exercise correlates with group 1 of the second phishing exercise. This confirms the use of group 1 (circular) as the baseline. Contrasting both exercises, we notice an improvement related to submitted data, reporting, and resilience, as shown in Table~\ref{tab:comparison-exercises}. Between the groups, differences can be noticed. The error rate of group 1 is almost doubled in comparison to the error rates of both other groups. Nevertheless, the rate improved in the second simulated phishing exercise. In addition, the reporting rate doubled in group 2 and tripled in group 3.

\begin{table}[H]
	\centering
	\caption{Comparison of the two simulated phishing exercises}\label{tab:comparison-exercises}
	\begin{tabular}{lcc}
		\toprule
		\textbf{KPIs} & \textbf{Exercise 1} & \textbf{Exercise 2} \\ \midrule
		Email opened & 6\% & 12\%  \\ 
		Clicked link & 9\% & 2\%  \\ 
		Submitted data & 17\% & 5\%  \\ 
		Reporting rate & 13\% & 23\%  \\ 
		Resilience factor & -1.5 & 4.6  \\ \bottomrule
	\end{tabular}
\end{table}

\subsection{Post-Study}
\label{sec:post}

Directly after the second simulated phishing exercise, a post-study in the form of a survey was conducted in the SME. Out of 93 employees, 21 participated. Although this rate seems low, it is higher than in the pre-study. Four employees, who did not receive any awareness training as part of the involved group, 
participated as well in the survey. Most participants were members of group 3, whereas group 1 had the lowest participation rate. Compared to the pre-study, the self-evaluation of the technical knowledge was slightly higher. In both studies, technical-savvy was rated higher than knowledge about phishing. Only two participants knew of any other awareness training campaign conducted. On average, the participants were able to name around four phishing characteristics. Also, the quality of the input (0: worst, 2: best), which depends on the number of stated characteristics per person, is similar to the pre-study. When focusing on the main characteristics of the sender and link by applying weights, more participants in the post-study recognized them. Group 3 performed slightly better than the average (see Table~\ref{tab:comparison-study}).

\begin{table}[!htpb]
	\centering
	\caption{Comparison of the stated characteristics of phishing between both studies}\label{tab:comparison-study}
	\begin{tabular}{lccc}
		\toprule
		\textbf{Avg. characteristics} & \textbf{Pre-study} & \textbf{Post-study} & \textbf{Group 3}\\ \midrule
		Ex. weight & 4.14 & 4.43 & 4.58 \\
		Quality ex. weight & 1.36 & 1.38 & 1.42  \\ 
		With weight & 5.53 & 6.14 & 6.33 \\ 
		Quality with weight & 1.5 & 1.67 & 1.67  \\ 
		\bottomrule
	\end{tabular}
\end{table}

\begin{figure*}[!htpb]
	\centering
	\includegraphics[width=0.9\textwidth]{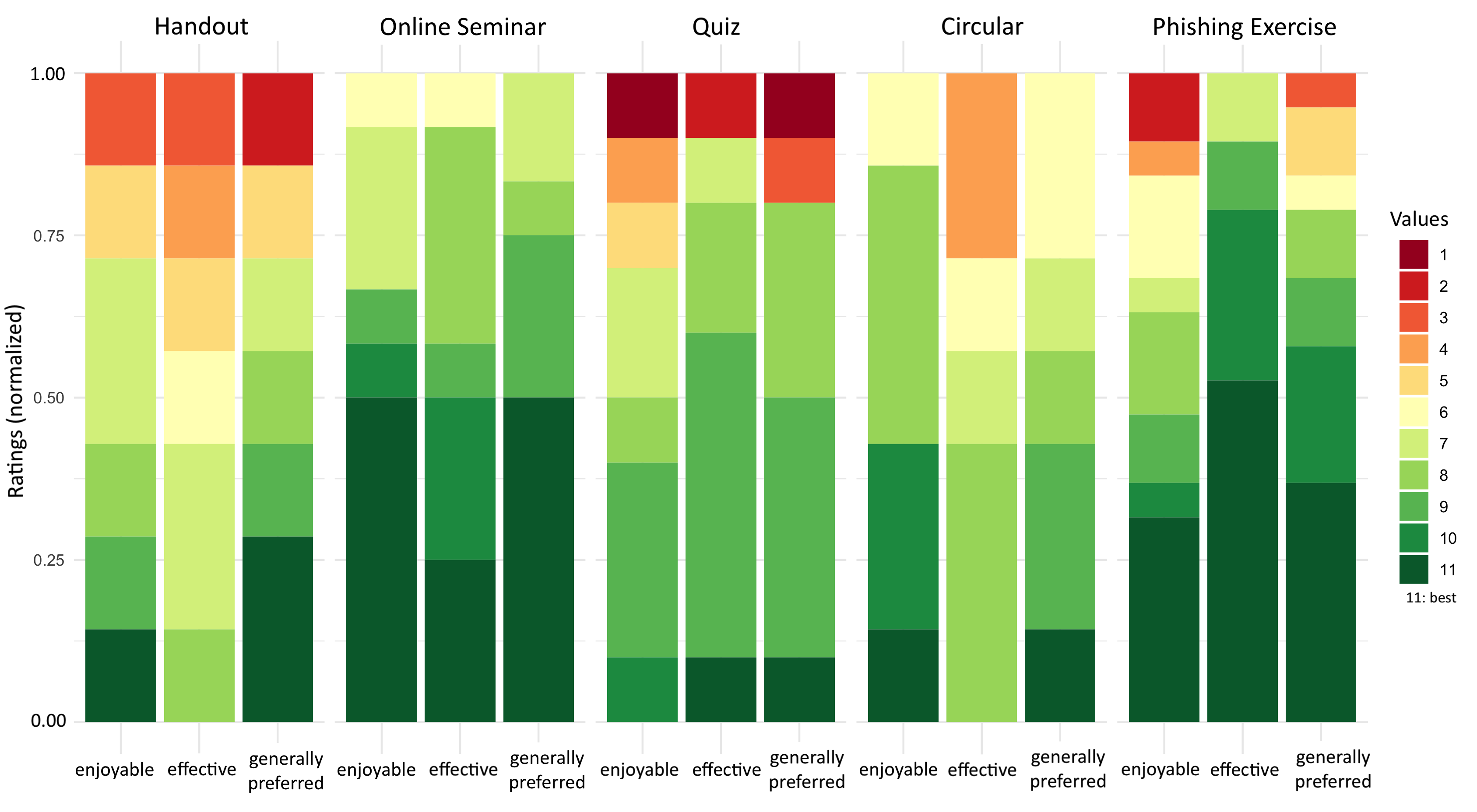}\\ 
	\caption{Comparison of preferences between different awareness trainings in post-study}\label{fig:post-study} 
	\Description[Comparison of preferences between different awareness trainings in post-study]{The comparison of preferences between different awareness trainings in the post-study shows that the online seminar is favored, followed by circular and phishing exercise, which is rated effective but not enjoyable.} 
\end{figure*}

Similar to the pre-study, the effectiveness and joyfulness of awareness training campaigns are evaluated by the participants. Most factors have similar tendencies and vary between 0.23 and 0.42. On average, the participants rate the awareness training positively. To evaluate the applied campaigns, only the participants are included. The results can be seen in Figure~\ref{fig:post-study}. The online training was significantly rated positively. No negative result was stated, and on average, this method was preferred. Especially the participants in group 3 chose their training. Regarding effectiveness, the simulated phishing exercise got the highest ratings. Although the effectiveness was graded as good, the preferability was estimated as less enjoyable. Based on the stated requirements of awareness training, no reason for this decision can be found. Nevertheless, the simulated phishing exercise was suggested as a combined method. Hence, we assume that it offers a trade-off between effectiveness and joyfulness. The quiz and circular received partly bad ratings. Both methods are asynchronous and might disappear from the mind due to the daily work routine. Interestingly, the circular is rated better than the handout. This might be due to the time factor, as the handout is more detailed and the employees first have to find it in the handbook. The effectiveness of both is seen as lower, as regularity (one requirement) is currently not given. In comparison to the pre-study, less diverse requirements for awareness training campaigns were stated, while ease gained importance.

When analyzing the combination of different awareness training methods, it is noticeable that the participants of the post-study mainly named the awareness training campaigns that were given during the assessment. The combination of phishing exercises, online training, and a quiz is slightly preferred. In addition, the participants requested further information on eight IT security topics.

\section{Discussion}
\label{sec:discussion}

As there is no consensus on which type of awareness training is best for employees, we designed and implemented three different awareness training campaigns in an SME. We verified the results based on two simulated phishing exercises, which were accompanied by surveys. In order to measure the success rate of the awareness training campaigns, different measures were observed. The main focus was on the different awareness training methods, as verified by the exercises. Hereby, we noticed differences between the three groups. The results enable the use of KPIs. Nevertheless, success depends on participation, and a distinct causality is difficult to obtain. To raise awareness and measure success, the level of participation has to be high. As not all participants also completed the surveys, the validity is reduced. However, some improvements were noticed. To increase resilience in the long-term, regular awareness training is required. This might include variants such as multi-factor authentication, the Open Authentication (OAuth) protocol, and chats. The recommendations for the best interval range from one month to half a year. Here, the internal support team has to balance the need for updated information with its efforts. At least, new employees should receive awareness training as soon as possible.

This study assumes that the success of awareness training depends on its acceptance by employees. If training campaigns are perceived as comfortable, then this results in a higher success rate. This assumption is confirmed by the second simulated phishing campaign and the post-study. The awareness training with the highest attendance rate, which is also seen as the preferred one (online training), had the greatest success. Based on both surveys, the training method should be straightforward, time-saving, pleasant, and informative. It should be easy to integrate into the work routine and offer additional value. The handout was rated worst as it is informative, but not time-saving. The online training was seen as more pleasant and effective than the quiz, although it takes more time. Especially the simulated phishing exercise was criticized in related work (see Section~\ref{sec:relatedwork}) as it may raise doubts and receive less acceptance. In the post-study, the employees rate the phishing exercise as a trade-off between effectiveness and joyfulness. In general, it was seen as rather enjoyable, but very effective. We are unsure if the same results could be gained from other variants, such as embedded training where employees receive education after falling for a simulated phishing email. Hence, motivating the employees and having a contact person in case of questions might lead to better results. The percentage of attendance and survey findings may indicate how well a awareness campaign performs.

One might argue that the results are due to the differences in depth and time of exposure to the awareness campaigns. However, we tried to convey the same information throughout the three awareness campaigns. If participants read the guidelines in the handbook, then this would even take longer.

Furthermore, different types of KPIs (see Table~\ref{tab:comparison-group}) lead to various possibilities of comparing exercises and group performance. In particular, the amount of people who opened an email without clicking on a link or submitting data is less meaningful than the other indicator values. As shown in Figure~\ref{fig:gophish_events} the measurement of this KPI is even more time-consuming than others and hardly scalable to a larger group. To continuously monitor and improve the performance, suitable KPIs and time frames have to be chosen. At least, the KPI regarding submitted data should be measured on a regular basis as this value has the most relevance to awareness. 

\section{Related Work}
\label{sec:relatedwork}

Several technical measures try to protect against phishing, as summarized by Patil and Arra~\cite{9753912}. Spam filters may classify emails as spam based on different characteristics, such as header, subject, content, URLs, and anomalies~\cite{8529061}. They often use machine learning, but may be circumvented~\cite{10.1145/3407023.3407079}. In addition, browser plugins, antivirus systems, and intrusion detection systems may notice phishing websites~\cite{10.1145/3299815.3314437}. We applied SPF in the phishing exercises. Furthermore, domain keys identified mail (DKIM) and domain-based authentication, reporting and conformance (DMARC) can be used to increase the validity of the sender's email address~\cite{10.1145/3485983.3494868}. This though does not prevent phishing. The authenticity of the sender can be verified by the use of digital signatures, e.\,g., pretty good privacy (PGP) or secure/multipurpose Internet mail extensions (S/MIME)~\cite{10008382}. On the other hand, they establish a false security~\cite{10.1145/3372297.3417878} and credibility~\cite{icissp21}. The introduction of multi-factor authentication can increase the level of security. However, adversaries may circumvent specific configurations and factors~\cite{10.1145/3440712,10.1145/3433210.3453084}.

This shows that additional organizational measures are required. Typically, the technical measures are assisted by policies (e.\,g., password policies, the need-to-know principle, and audits) as part of the ISMS. To train employees, different awareness training methods exist, which have various advantages and disadvantages. For example, Ghafir et al.~\cite{ghafir} explain that the speed of the training may not suit every participant. As our training was recorded, participants could watch it at different rates. Higashino et al.~\cite{8714691} introduce a simulated phishing exercise, which saves the information to a trainer's computer. Blythe et al.~\cite{10.1007/978-3-030-50309-3_6} analyze rewards and sanctions after simulated phishing training with the result of no influence. Sutter et al.~\cite{9893815} assess different variants of embedded training. Volkamer et al.~\cite{volkamer} raise concerns related to security, law, and the human factor when conducting phishing campaigns. As we did not pinpoint individual participants and applied a transparent process, the phishing awareness campaigns were received positively. After solving the quiz PHISHY, the participants better recognized phishing emails, according to Gokul et al.~\cite{phishy}. In contrast, newsletters are easy to receive, but similar information and structure over time reduce the attention of participants as shown by Sendelbach et al.~\cite{fatigue}. Since group 1 received an email with information not for the first time, this result is likely to be confirmed.

Although several awareness campaigns are proposed, almost no study actually compares them. In addition, the iteration for training sessions is unclear. Innab et al.~\cite{8593144} contrast the phishing awareness of governmental and private organizations in Riyadh without describing the specific awareness campaigns. Carella et al.~\cite{8258485} evaluate no training with document-based training and in-class training. The authors conclude that document-based training is the most effective, which is contrary to our results. Ikhsan et al.~\cite{ikhsan} apply phishing simulation and a questionnaire, without comparing them. The authors also show that if the KPIs are high, then there is a need for action. 
Reinheimer et al.~\cite{10.5555/3488905.3488920} recommend conducting training at least every six months for a constant awareness rate. Grimes even suggests a monthly rate. Based on our study, the KPIs can be used to dynamically adjust the iterations and balance the need for updated information with the effort required for it.

\section{Conclusion and Outlook}
\label{sec:conclusion}

Phishing is a major cyber threat to organizations. As technical measures do not work 100\%, awareness is required. However, there is little validation of awareness measures. Consequently, we examined how awareness concepts can be successfully implemented and validated. For this purpose, we outlined the methodology of our study, consisting of a pre-study, the main study with simulated phishing exercises and awareness campaigns, and a post-study. The concept detailed the methodology by selecting, for example, the tool for the phishing exercise and the three executed awareness training campaigns. The results of the phishing study showed that pleasant awareness campaigns (see Figure~\ref{fig:post-study}) lead to a higher success rate in detecting phishing attacks (see Table~\ref{tab:comparison-group}). In addition, training campaigns have to be easily integrated into the daily work routine. After discussing our results, we contrasted them with related work. 

Over all, the online training (group 3) has produced the best results in our conducted studies. Even in contrast to a well-designed quiz (group 2) which was quite shorter in duration, it performed better. Further comparisons of awareness methods should be carried out to verify and deepen the research by refining the concept of the methods and increasing the number of participants.

In the future, we want to further reduce data protection concerns by improving the phishing tool. To increase the resilience of the employees, we plan to vary the phishing emails and include other social engineering variants. This could be done in a long-term study to see variations over time. Similar exercises in other organizations could reveal similarities and differences. Laslty, examining the reasons for the error rate may give us more insight into how to improve individual performance.

\begin{acks}
We thank LMU's Statistical Consulting Unit (StaBLab) for their support to prepare and analyze the data.
\end{acks}

\bibliographystyle{ACM-Reference-Format}
\bibliography{phishing-awareness}

\end{document}